\documentclass[12pt]{iopart}

\usepackage{iopams,graphicx,amssymb}
\eqnobysec

\begin{document}

\title[Critical behaviour of the $O(n)$-$\phi^{4}$ model with an
antisymmetric order parameter]
{Critical behaviour of the $O(n)$-$\phi^{4}$ model with an
antisymmetric tensor order parameter}

\author{N. V. Antonov, M. V. Kompaniets and N. M. Lebedev}

\address{Department of Theoretical Physics, St.~Petersburg University,
Uljanovskaja 1, St.~Petersburg, Petrodvorez, 198504 Russia}

\ead{nikolai.antonov@pobox.spbu.ru, mkompan@gmail.com}

\begin{abstract}
Critical behaviour of the $O(n)$-symmetric $\phi^{4}$-model with
an antisymmetric tensor order parameter is studied by means of the
field-theoretic renormalization group (RG) in the leading order of
the $\varepsilon=4-d$-expansion (one-loop approximation). For
$n=2$ and $3$ the model is equivalent to the scalar and the
$O(3)$-symmetric vector models, for $n\ge4$ it involves two
independent interaction terms and two coupling constants. It is
shown that for $n>4$ the RG equations have no infrared (IR)
attractive fixed points and their solutions (RG flows) leave the
stability region of the model. This means that fluctuations of the
order parameter change the nature of the phase transition from the
second-order type (suggested by the mean-field theory)
to the first-order one. For $n=4$, the IR attractive fixed point
exists and the IR behaviour is non-universal: if the coupling
constants belong to the basin of attraction for the IR point, the
phase transition is of the second order and the IR critical
scaling regime realizes. The corresponding critical exponents
$\nu$ and $\eta$ are presented in the order $\varepsilon$ and
$\varepsilon^{2}$, respectively. Otherwise the RG flows pass
outside the stability region and the first-order transition takes
place.

\end{abstract}


\section{Introduction} \label{sec:Intro}

Numerous physical systems reveal interesting singular behaviour in the
vicinity of their critical points. Their thermodynamical and correlation
functions exhibit scaling behaviour with universal critical dimensions:
according to the prevailing belief, they depend only on few global
characteristics of the system, like symmetry or dimension; see, e.g.,
the monographs \cite{Brout}.
Preliminary analysis of the critical behaviour is usually performed within
the framework of the phenomenological Landau theory, where the free energy
of the system in question is written in the simplest form dictated by the
symmetry \cite{LL}. That approach predicts the type of
the phase transition (first- or second-order one) but gives only
approximate ``mean-field'' values for the critical exponents.
More refined fluctuation theory applies Landau's idea to the effective
Hamiltonian of the system, which is written in a form similar to a certain
Euclidean field theoretic model \cite{LL}. Thus the further analysis of the
problem calls for the field theoretic techniques.

The powerful and quantitative theory of the critical behaviour is provided
by the field theoretic renormalization group (RG); see the monographs
\cite{Zinn,Book3,Klein} and the literature cited therein. In the RG approach,
possible types of critical behaviour (universality classes) are associated
with infrared (IR) attractive fixed points of renormalizable field theoretic
models. Most typical phase transitions (liquid-vapour systems, binary alloys,
ferro- and antiferromagnets) belong to the universality class of the
$O(n)$-symmetric model with quartic interaction (Euclidean $\phi^{4}$-model)
with an $n$-component order parameter. Another important example is provided
by the $U(n)$-symmetric $\phi^{4}$-model with a complex order parameter.
That model, describing transitions in quantum gases and liquids, is in fact
equivalent to the $O(2n)$-symmetric real case.

In agreement with the Landau theory that predicts a second-order transition
for such systems, the RG analysis establishes therein the existence of
nontrivial IR attractive fixed point in the physical range of parameters,
and hence the existence of IR scaling behaviour. Its universal
characteristics depend only on $n$ and $d$, the dimension of the
system, and can be systematically calculated as expansions in
$\varepsilon=4-d$, the deviation of the spatial dimension from its upper
critical value $d=4$ \cite{Zinn,Book3,Klein}.

In many cases, however, description by the aforementioned, relatively
simple, models appears inadequate, and one has to consider more
sophisticated symmetries or more complex types of the order parameter with
tensor or matrix nature. Not an exhaustive list of such phenomena
includes phase transitions in systems with nontrivial crystallographic
symmetry or randomly distributed impurities (see the monograph \cite{PPV}
for a general review and references), various transitions in liquid crystals
\cite{ZR,DeG,Priest,Shal,Leo1}, transitions between different superfluid
phases in He$^{3}$ \cite{Mineev,He,He3} and in the neutron liquid in neutron
stars \cite{NS,NS1}, transition to superconductive state in systems with
higher spins \cite{NKH}, models of Laplacian growth with multifractal
properties \cite{MF}, and so on.

As a rule, the corresponding field-theoretic models involve several types
of interaction terms and hence several coupling constants (charges). The
corresponding RG equations can have several fixed points with different
attractive properties \cite{Priest,Shal,He,He3,NS,NS1,NKH,Vicari}. This
can lead to a very complicated pattern of the corresponding RG flows
(solutions of the RG equations for the invariant charges) in the space of
model parameters.

Although some general statement (the so-called $\eta$ conjecture) can be
formulated for the IR attractive fixed points in models with an $n$-component
vector order parameter \cite{Vicari}, the very existence of such points
is not a necessary feature of the $\phi^{4}$-models. Furthermore,
even in the presence of IR attractive points, the RG flow can pass outside
the natural region of parameters, determined by the stability of the system,
the situation usually interpreted as a first-order phase transition.
It may also go to infinity, the situation that lies beyond the scope of the
perturbation theory. As a result, the predictions of the plain Landau theory
can be essentially corrected.

In this connection, the Ginzburg-Landau model of superconductivity \cite{LL}
is worth recalling: the one-loop analysis of the corresponding field
theoretic model (actually, the electrodynamics of a charged scalar field)
shows that is has admissible fixed point only for very large $n$ \cite{Sc1}.
The situation, however, is not completely clear: two-loop calculations with
an appropriate resummation procedure suggest that the attractive point ``has
a chance to exist'' \cite{Sc2}. The non-perturbative analysis of Ref.
\cite{Leo2} also favors the second-order transition.

In a sense, opposite examples are provided by the model with a
symmetric tensor order parameter and by the Potts model: according to the
Landau theory, existence of a cubic term excludes the possibility of the
second-order transition. On the contrary, exact two-dimensional results,
numerical simulations and RG analysis suggest that for small $n$, the phase
transition is of the second order \cite{Priest,Amit}.

In this paper we apply the field theoretic RG to the $O(n)$-symmetric
$\phi^{4}$-model of the real $n$-th rank tensor order parameter. This
model can be relevant in the analysis of transitions between the nematic,
cholesteric and blue phases in liquid crystals \cite{Blue1,Blue},
transitions to ferroelastic state in solids \cite{FerEl,FerEl1} and
transitions to superconductive state in systems with higher spins \cite{NKH}.
Our main motivation, however, is more theoretical, and in order to simplify
and to sharpen the problem, we consider the case of a purely antisymmetric
tensor. In comparison to the general $n$-th rank tensor case, this reduces
the model to the two-charge problem, which makes the results more visible.
The model is probably the simplest one with a non-vector order parameter,
but remains a multicharge one and, as we will show, demonstrates the features
typical of more realistic and complex situations listed above. In this
connection it is also important that the cubic invariant for the purely
antisymmetric tensor vanishes identically, so that the Landau theory, as
conventionally applied, predicts a second-order transition, and we do not
face the contradiction, mentioned above for the symmetric tensor case.

The plan of the paper is as follows. In the next three sections we formulate
the model, give the corresponding Feynman rules, perform the ultraviolet (UV)
renormalization and give the explicit leading-order expressions for the
renormalization constants. In section~\ref{sec:RGE} we present the RG
equations for the renormalized Green functions and give the leading-order
expressions for their coefficients ($\beta$-functions and anomalous dimensions).
In section~\ref{sec:FP} we analyze the fixed points of the RG equations for
the invariant coupling constants.

It turns out that existence of an IR attractive fixed point in the physical
range of parameters is an exception rather than a rule. Such points exist
for the special cases $n=2$ and $n=3$, when our model becomes equivalent to
the scalar and the $O(3)$-symmetric vector models, respectively, and thus
is in fact a single-charge one. For general $n$, admissible fixed points
exist only in a certain single-charge special case of the model, which
appears multiplicatively renormalizable in itself (that is, closed with
respect to the renormalization procedure) and thus can be studied as a
separate internally consistent model (in the full-scale two-charge model,
such fixed points are saddle points).

The only admissible fixed point in the full-scale problem exists for $n=4$,
that is, for the minimal possible value of the rank where the model is a
genuine two-charge one (for larger $n$ that point becomes complex). Its
existence means that all the Green functions of the model in the IR range
can demonstrate self-similar (scaling) behaviour. The corresponding critical
exponents $\eta$ and $\nu$ are given in section~\ref{sec:Conc} in the leading
order of the $\varepsilon$-expansion. However, for a multicharge model, even
when an IR attractive fixed point is present, not every ``RG flow'' (solution
of the RG equations for the invariant charges) approaches it in the IR
asymptotic range: it can first pass outside the region of stability
(an indication of the first-order transition) or go to infinity (then no
definitive conclusions can be drawn within the perturbation theory).

The main conclusion is that the account of fluctuations can change the
character of the phase transition for the antisymmetric order parameter
from the second-order to the first-order type; this is a possible
behaviour for $n=4$ and the only possible one for all $n>4$.

\section{The model} \label{sec:Model}

We study a model of a real antisymmetric $n$-th rank tensor field
$\phi = \phi_{ik} ({\bf x})$ (so that $\phi_{ik}=-\phi_{ki}$
and $i,k=1,\dots,n$) in the Euclidean $d$-dimensional ${\bf x}$ space.
The action functional has the form
\begin{equation}
{\cal S}(\phi)= {\cal S}_{0}(\phi) + V(\phi)
\label{action}
\end{equation}
with the free part
\begin{equation}
{\cal S}_{0}(\phi) = \frac{1}{2} \tr \{\phi(-\partial^{2}+\tau_{0})
\phi\}
\label{free}
\end{equation}
and the interaction term with the two independent quartic structures
\begin{equation}
V(\phi) = V_{1}(\phi) + V_{2}(\phi) =
- \frac{g_{10}}{4!} \{\tr(\phi^{2})\}^2
- \frac{g_{20}}{4!} \tr (\phi^{4}).
\label{V}
\end{equation}
Here (and in analogous formulas below) integration over the
$d$-dimensional ${\bf x}$ space is implied; $\partial^{2}$ is the Laplace
operator, $\tau_{0}$ is the deviation of the temperature (or its analog)
from the critical value and $g_{10}$, $g_{20}$ are the coupling constants.
In the detailed notation
\[  {\cal S}_{0}(\phi)= - \frac{1}{2} \int d{\bf x}
\sum_{i,k=1}^{n} \phi_{ik}({\bf x})
(-\partial^{2}+\tau_{0}) \phi_{ik}({\bf x}) \]
and similarly for $V(\phi)$.
The cubic term $\tr(\phi^{3})$ vanishes due to the antisymmetry of
$\phi$ and does not appear in the interaction.

Correlation functions (Green functions) of the model are given by the
functional averages with weight $\exp{\cal S}(\phi)$. The action
(\ref{action})--(\ref{V}) is invariant with respect to the transformation
$\phi \to {\cal O} \phi {\cal O}^{\dagger}$, where ${\cal O}\in O(n)$
is an $n$-th rank orthogonal matrix (note that the antisymmetry property
is preserved by this transformation).

The stability of the model requires that the interaction term (\ref{V})
be negative for all values of $\phi$. One can check that the condition
$V(\phi)<0$ imposes the following restrictions on the coupling constants:
\begin{equation}
2 g_{10} + g_{20} >0, \quad  n g_{10} + g_{20} >0
\label{stab1}
\end{equation}
for even values of $n$ and
\begin{equation}
2 g_{10} + g_{20} >0, \quad  (n-1) g_{10} + g_{20} >0
\label{stab2}
\end{equation}
for $n$ odd.

For $n=2$ and $n=3$ the model (\ref{action})--(\ref{V}) reduces to the
well-known cases: the single-component $\phi^{4}$-model and the
$O(3)$-invariant vector model, respectively. The correspondence can be
established by means of the transformations
$\phi_{ik}=\varepsilon_{ik} \phi$ for $n=2$ and
$\phi_{ik}=\varepsilon_{ikl} \phi_{l}$ for $n=3$,
where the both $\varepsilon$'s are fully antisymmetric tensors
with normalization $\varepsilon_{12}=\varepsilon_{123}=+1$.
Then the both structures in $V(\phi)$ become identical to $\phi^{4}$
for $n=2$ and $(\phi_{l}\phi_{l})^{2}$ for $n=3$, and the only
remaining coupling constant is the combination $g_{0}=2g_{10}+g_{20}$.
In the both cases, the stability conditions (\ref{stab1}), (\ref{stab2})
reduce to the single inequality $g_{0}>0$.

\section{Diagrammatic techniques} \label{sec:TV}

The Feynman diagrammatic techniques for the model (\ref{action})--(\ref{V})
is derived in a standard fashion; see e.g. \cite{Zinn}--\cite{Klein}.
In the momentum (Fourier) representation the bare propagator, determined
by the free action (\ref{free}), has the form
\begin{equation}
\langle \phi_{ik} \phi_{lm} \rangle_{0} =
\frac{J_{ik;lm}} {(p^{2}+\tau_{0})},
\label{prop}
\end{equation}
where $p=|{\bf p}|$ is the wave number. The tensor
\begin{equation}
J_{ik;lm} = \frac{1}{2} \left(\delta_{il} \delta_{km} -
\delta_{im} \delta_{kl}\right),
\label{J}
\end{equation}
built of the Kronecker $\delta$ symbols, is antisymmetric with respect
to the transpositions of its indices $i \leftrightarrow k$ and
$l \leftrightarrow m$, and symmetric with respect
to the transposition of the pairs $ik \leftrightarrow lm$.
It plays the part of the unit operation on the space of antisymmetric
tensors in the sense that $J_{ik;lm}\phi_{lm}=\phi_{ik}$ and
$J_{ik;lm}J_{lm;js}=J_{ik;js}$. Its ``trace'' with respect to the pairs
of indices $J_{ik;ik}=n(n-1)/2$ gives the number of independent components
of an $n$-th rank antisymmetric tensor.

The interactions $V_{1,2}(\phi)$ in (\ref{V}) correspond to the quartic
vertices with the vertex factors $(-g_{10}) V^{(1)}_{ab;cd;ef;mn}$
and $(-g_{20}) V^{(2)}_{ab;cd;ef;mn}$, where the tensors
\begin{equation}
V^{(1)}_{ab;cd;ef;mn} = \frac{1}{3} \left( J_{ab;cd}J_{ef;mn} +
J_{ab;ef}J_{cd;mn} + J_{ab;mn}J_{cd;ef} \right)
\label{V1}
\end{equation}
and
\begin{eqnarray}
V^{(2)}_{ab;cd;ef;mn} = \frac{1}{6} &{}& \biggl(
J_{ab;ij} J_{cd;jk} J_{ef;kp} J_{mn;pi} +
J_{ab;ij} J_{cd;jk} J_{mn;kp} J_{ef;pi} +
\nonumber \\
&+& J_{ab;ij} J_{ef;jk} J_{mn;kp} J_{cd;pi} +
    J_{ab;ij} J_{ef;jk} J_{cd;kp} J_{mn;pi} +
\nonumber \\
&+& J_{ab;ij} J_{mn;jk} J_{ef;kp} J_{cd;pi} +
    J_{ab;ij} J_{mn;jk} J_{cd;kp} J_{ef;pi} \biggr)
\label{V2}
\end{eqnarray}
are defined such that
\[ V^{(1)}_{ab;cd;ef;mn} \phi_{ab}\phi_{cd}\phi_{ef}\phi_{mn} =
\{tr (\phi^{2})\}^{2} \]
and
\[ V^{(2)}_{ab;cd;ef;mn} \phi_{ab}\phi_{cd}\phi_{ef}\phi_{mn} =
tr (\phi^{4}), \]
and such that they are antisymmetric with respect to the transpositions
of the indices $a \leftrightarrow b$, $c \leftrightarrow d$ and so on,
and symmetric with respect to the transpositions of the pairs
$ab \leftrightarrow cd$, $ab \leftrightarrow mn$ and so on.

Thus any diagram of our model is represented as a product of two factors:
the corresponding diagram for the single-component $\phi^{4}$-model with
the corresponding symmetry coefficient and the additional $n$-dependent
factor stemming from the contractions of the tensors in the propagators
(\ref{prop}) and vertices (\ref{V1}), (\ref{V2}).

\section{UV renormalization} \label{sec:UV}

The analysis of renormalizability of the model (\ref{action})--(\ref{V}) is
very similar to the case of the single-component $\phi^{4}$-model; see e.g.
\cite{Zinn}--\cite{Klein}. The model is logarithmic (the coupling constants
$g_{10}$, $g_{20}$ are dimensionless) for $d=4$. In the dimensional
regularization, the UV divergences have the form of the poles
in $\varepsilon=4-d$, deviation of the dimension of space from its upper
critical value $d=4$. Standard analysis, based on the dimensionality and
symmetry considerations, shows that superficial UV divergences, whose
elimination requires counterterms, are present only in the 1-irreducible
Green functions $\langle\phi\phi\rangle$ and $\langle\phi\phi\phi\phi\rangle$.
The needed counterterms have the same forms as the terms already present in
the action and can therefore be reproduced by the multiplicative
renormalization of the field and the model parameters.

The corresponding renormalized action has the form
\begin{equation}
{\cal S}_{R}(\phi)=
\frac{1}{2} \tr \{\phi(-Z_{1}\partial^{2}+ Z_{2}\tau) \phi\}
- \frac{g_{1}\mu^{\varepsilon}}{4!} Z_{3} \{\tr(\phi^{2})\}^2
- \frac{g_{2}\mu^{\varepsilon}}{4!} Z_{4} \tr (\phi^{4}).
\label{actionR}
\end{equation}
Here $\tau$, $g_{1}$ and $g_{2}$ are renormalized analogs of the bare
parameters (with the subscripts ``o'') and $\mu$ is the reference mass
scale (additional arbitrary parameter of the renormalized theory).
Expression (\ref{actionR}) can be reproduced by the
multiplicative renormalization of the field $\phi \to \phi Z_{\phi}$
and the parameters:
\begin{eqnarray}
\tau_{0} = \tau Z_{\tau},  \quad
g_{01} = g_{1} \mu^{\varepsilon} Z_{g_{1}}, \quad
g_{02} = g_{2} \mu^{\varepsilon} Z_{g_{2}},
\label{Mult}
\end{eqnarray}
so that
\begin{equation}
Z_{1} = Z_{\phi}^{2}, \quad Z_{2} = Z_{\tau}Z_{\phi}^{2},  \quad
Z_{3} = Z_{g_{1}}Z_{\phi}^{4} , \quad
Z_{4} = Z_{g_{2}}Z_{\phi}^{4}.
\label{alg}
\end{equation}
We use the minimal subtraction (MS) scheme, where the all renormalization
constants $Z_{i}$ have the forms ``1+\, only poles in $\varepsilon$,''
\begin{eqnarray}
Z_{i}=1 + \sum_{n=1}^{\infty} A_{in}(g_{1,2})\, \varepsilon^{-n},
\label{MS}
\end{eqnarray}
with
the coefficients depending only on the completely dimensionless renormalized
couplings $g_{1,2}$. The constants $Z_{1,2}$ and $Z_{3,4}$ are calculated
directly from the two-point and four-point 1-irreducible Green functions,
respectively, then the constants in (\ref{Mult}) are found from the relations
(\ref{alg}).

The explicit one-loop calculation gives
\begin{eqnarray}
Z_{2}= 1+ \frac{1}{12\varepsilon} \{(n^{2}-n+4) g_{1} + (2n-1)g_{2}\},
\label{Z2}
\end{eqnarray}
\begin{eqnarray}
Z_{3}= 1+ \frac{1}{12\varepsilon} \{(n^{2}-n+16) g_{1} + 2(2n-1)g_{2} +
3 g_{1}^{2}/g_{2}\},
\label{Z3}
\end{eqnarray}
\begin{eqnarray}
Z_{4}= 1+ \frac{1}{12\varepsilon} \{24 g_{1} + (2n-1)g_{2} \},
\label{Z4}
\end{eqnarray}
with the corrections of the order $g_{1,2}^{2}$. In order to simplify
the coefficients, here and below we pass to the new couplings:
$g_{1,2}\to g_{1,2}/(8\pi^2)$.

One can show that the model (\ref{action})--(\ref{V}) with $g_{20}=0$ is
multiplicatively renormalizable in itself: the interaction $V_{1}$ alone
does not generate the structure $V_{2}$ in the counterterms. On the contrary,
the model with $g_{10}=0$ is not closed with respect to renormalization: the
interaction $V_{2}$ gives rise to the both structures $V_{1,2}$. This leads
to the appearance of the coupling $g_{2}$ in the denominator of the
one-loop expression (\ref{Z3}).

Like in the ordinary $\varphi^{4}$-model, the nontrivial contributions
to the constant $Z_{1}$ appear only in the two-loop order:
\begin{eqnarray}
Z_{1}= 1 - \frac{1}{2\cdot24^{2}\varepsilon}
\{(n^{2}-n+4) (4g_{1}^{2} + g_{2}^{2})
+8(2n-1)g_{1}g_{2}\},
\label{Z1}
\end{eqnarray}
with the corrections of the order $g_{1,2}^{3}$.

\section{RG equations and RG functions} \label{sec:RGE}

The RG equations for the renormalized Green functions in a multiplicatively
renormalizable model are derived in a standard fashion; see e.g.
\cite{Book3}. In the model (\ref{actionR}) the RG equation for the
renormalized $n$-point function $W^{R}_{n}$  has the form:
\begin{eqnarray}
\{{\cal D}_{\mu} + \beta_{1} \partial_{g_{1}} + \beta_{2} \partial_{g_{2}}
- \gamma_{\tau}{\cal D}_{\tau} - n \gamma_{\varphi}\} W^{R}_{n} =0.
\label{RGE}
\end{eqnarray}
where ${\cal D}_{x}\equiv x\partial_{x}$ for any variable $x$.

The RG functions ($\beta$-functions for the coupling constants and anomalous
dimensions $\gamma$) are defined by the relations
\begin{equation}
\gamma_{i}\equiv \widetilde {\cal D}_{\mu}\ln Z_{i}\quad {\rm for\ any} \
Z_{i},
\label{RGF}
\end{equation}
where $\widetilde {\cal D}_{\mu}$ is the operation ${\cal D}_{\mu}$ at
fixed bare parameters and
\begin{equation}
\beta_{i} \equiv \widetilde {\cal D}_{\mu} g_{i} =
g_{i}\,[-\varepsilon-\gamma_{g_{i}}], \quad i=1,2,
\label{betagw}
\end{equation}
where the second equalities come from the definitions and the relations
(\ref{Mult}).  In the MS scheme the anomalous dimensions depend only on
the couplings $g_{1,2}$ and are given by simple expressions
\begin{equation}
\gamma_{i} = - \left( {\cal D}_{g_{1}} + {\cal D}_{g_{2}} \right)
A_{i1}(g_{1,2}),
\label{RG1}
\end{equation}
where $A_{i1}$ is the coefficient in the first-order pole in $\varepsilon$
in (\ref{MS}). In our approximation $Z_{\tau}=Z_{2}$, $Z_{g_{1}}=Z_{3}$ and
$Z_{g_{2}}=Z_{4}$, and from equation (\ref{RG1}) and the explicit expressions
(\ref{Z2})--(\ref{Z1}) we obtain:
\begin{eqnarray}
\beta_{1} = -\varepsilon g_{1} + \frac{1}{12} (n^{2}-n+16) g_{1}^{2} +
\frac{1}{6} (2n-1)g_{1}g_{2} +\frac{1}{4} g_{2}^{2} ,
\nonumber \\
\beta_{2}=-\varepsilon g_{2}+2 g_{1}g_{2} + \frac{1}{12} (2n-1)g_{2}^{2}
\label{beta1}
\end{eqnarray}
with the corrections of the order $g_{1,2}^{3}$ and
\begin{eqnarray}
\gamma_{\tau} = -  \frac{1}{12} \{(n^{2}-n+4) g_{1} + (2n-1)g_{2}\},
\nonumber \\
\gamma_{\varphi} = \frac{1}{2\cdot24^{2}}
\{(n^{2}-n+4) (4g_{1}^{2} + g_{2}^{2})
+8(2n-1)g_{1}g_{2}\},
\label{gamma1}
\end{eqnarray}
with the corrections of the order $g_{1,2}^{2}$.

\section{Fixed points and critical regimes} \label{sec:FP}

Possible asymptotic regimes of a renormalizable field theoretic model
are determined by the asymptotic behaviour of the system of ordinary
differential equations for the so-called invariant coupling constants
\begin{eqnarray}
{\cal D}_s \bar g_{i}(s,g) = \beta_{i} (\bar g), \quad
\bar g_{i}(1,g) = g_{i}.
\label{Odri}
\end{eqnarray}
Here $s=k/\mu$ is a nondimensionalized momentum, $g= \{g_{i}\}$ is the full
set of couplings and $\bar g_{i}(s,g)$ are the corresponding invariant
variables. As a rule, the IR ($s\to0$) and UV ($s\to\infty$) behaviour of
the Green functions is determined by fixed points $g_{i*}$ of the system
(\ref{Odri}). The coordinates of possible fixed points are found from the
requirement that all the $\beta$ functions vanish:
\begin{eqnarray}
\beta_{i} (g_{*}) =0.
\label{fp}
\end{eqnarray}
The type of a fixed point is determined by the matrix
\begin{equation}
\omega_{ik} = \partial\beta_{i}/\partial g_{k} |_{g=g_*}\, ,
\label{OmegaDef}
\end{equation}
which appears in the linearized version of the system (\ref{Odri}) near the
given point. For IR attractive fixed points (which we are interested in here)
the matrix $\omega$ is positive, {\it i.e.} the real parts of all
its eigenvalues $\omega_{i}$ are positive.

However, as already mentioned, for $n=2$ and $n=3$ our model reduces to the
single-charge scalar and $O(3)$-vector models, respectively. The only
coupling constant appearing in the Green functions is the combination
$g=2g_{1}+g_{2}$. From expressions (\ref{beta1}), (\ref{gamma1}) it is easily
checked that the corresponding $\beta$-function $\beta=2\beta_{1}+\beta_{2}$
and the anomalous dimensions $\gamma_{\varphi,\tau}$ depend on the only
parameter $g$ and coincide, up to the notation, with the known expressions
for the scalar and vector cases. An IR attractive fixed point with
$\beta(g_{*})=0$, $\beta'(g_{*})>0$ in the physical range $g_{*}>0$ exists
for $\varepsilon>0$.

For $n\ge4$ we have a genuine two-charge model. In renormalized perturbation
theory, the physical region of their values is given by the inequalities
(\ref{stab1}), (\ref{stab2}) with the replacement $g_{i0}\to g_{i}$:
\begin{eqnarray}
2 g_{1} + g_{2} >0, \quad  n g_{1} + g_{2} >0 \quad
{\rm for\ even\ } n
\nonumber \\
2 g_{1} + g_{2} >0, \quad  (n-1) g_{1} + g_{2} >0 \quad
{\rm for\ odd\ } n.
\label{stabR}
\end{eqnarray}
Analysis of the one-loop expressions (\ref{beta1}) reveals the following
fixed points:

1) Gaussian (free) fixed point $g_{1*}=g_{2*}=0$, UV attractive (IR
repulsive) for all $n$ with the eigenvalues $\omega_{1,2}=-\varepsilon$.

2) The point
\begin{equation}
g_{1*}= 12\varepsilon /(n^{2}-n+16), \quad g_{2*}=0.
\label{Fps}
\end{equation}
For all $n\ge4$ it lies in the physical region, but is a saddle point: the
eigenvalues $\omega_{1}=\varepsilon$ and
$\omega_{2}=-\varepsilon(n^{2}-n-8)/(n^{2}-n+16)$
are real and opposite in sign.

The relation $g_{2*}=0$ remains valid to all orders in $\varepsilon$. This
is a consequence of the fact that the model (\ref{action}) with $g_{2}=0$
is ``closed with respect to renormalization,'' see the end of
section~\ref{sec:UV}. For the single-charge model with the only interaction
$V_{1}$ in (\ref{action}) this point is IR attractive with the only
relevant eigenvalue $\omega_{1}=\varepsilon$.

3) Two fully nontrivial points with the both nonvanishing coordinates:
\begin{eqnarray}
g_{1*} = - 6\varepsilon \frac{ (4n^{2}-4n-143) \pm (2n-1)
\sqrt  {(-8n^{2}+8n+97)}} {(4n^{4}-8n^{3}-123n^{2}+127n+1696)},
\nonumber  \\
g_{2*} =  12\varepsilon \frac{ (2n-1)(n^{2}-n-20) \pm 12
\sqrt {(-8n^{2}+8n+97)}} {(4n^{4}-8n^{3}-123n^{2}+127n+1696)}.
\label{Sqrt}
\end{eqnarray}
For all $n\ge5$, however, these points are complex and thus cannot be
reached by the RG flow (\ref{Odri}) with real initial data. The only
exception is the case $n=4$, when the expressions (\ref{Sqrt}) become real
and take on the following simple forms:
\begin{eqnarray}
g_{1*} = 12\varepsilon/17, \quad  g_{2*} = - 12\varepsilon/17
\label{FP1}
\end{eqnarray}
for the plus sign in front of the square root in (\ref{Sqrt}) and
\begin{eqnarray}
g_{1*} = 9\varepsilon/11, \quad  g_{2*} = - 12\varepsilon/11
\label{FP2}
\end{eqnarray}
for the minus sign. The both points lie in the stability region
(\ref{stabR}). The first point is IR attractive with the eigenvalues
$\omega_{1}=\varepsilon$, $\omega_{2}=\varepsilon/17$, while the second
one is a saddle point with $\omega_{1}=\varepsilon$,
$\omega_{2}=-\varepsilon/11$.

\section{Discussion and Conclusion} \label{sec:Conc}

We conclude that for the genuine two-charge cases $n\ge4$ with the both
interactions $V_{1,2}$ an IR attractive fixed point in the stability
region exists only for $n=4$; in the one-loop approximation it is given
by the expression (\ref{FP1}). For the model with the only interaction
$V_{1}$ there is an IR attractive point (\ref{Fps}) for all $n$.

Existence of an IR attractive fixed point implies existence of scaling
behaviour for all the Green functions, described by the two main independent
critical exponents \cite{Zinn,Book3,Klein}
\begin{eqnarray}
\eta = 2\gamma^{*}_{\varphi}, \quad 1/\nu =2+ \gamma^{*}_{\tau},
\label{indi}
\end{eqnarray}
where $\gamma^{*}_i = \gamma_i (g_{1*},g_{2*})$ are the values of anomalous
dimensions (\ref{RGF}) at the fixed point in question.

For the fixed point (\ref{Fps}) from the explicit leading-order expressions
(\ref{gamma1}) one obtains
\begin{eqnarray}
\eta = \frac{(n^{2}-n+4)}{(n^{2}-n+16)^{2}}\,\varepsilon^{2}, \quad
1/\nu =2- \frac{(n^{2}-n+4)}{(n^{2}-n+16)}\,\varepsilon,
\label{indi1}
\end{eqnarray}
which for $n=4$ gives
\begin{eqnarray}
\eta = \frac{\varepsilon^{2}}{49}, \quad 1/\nu =2-\frac{4\varepsilon}{7}.
\label{indi14}
\end{eqnarray}
For the fixed point (\ref{FP1}) one has
\begin{eqnarray}
\eta = \frac{6\varepsilon^{2}}{289}, \quad 1/\nu =2-\frac{9\varepsilon}{17},
\label{indi2}
\end{eqnarray}
while for (\ref{FP2}) one obtains
\begin{eqnarray}
\eta = \frac{5\varepsilon^{2}}{242}, \quad 1/\nu =2-\frac{5\varepsilon}{11}.
\label{indi3}
\end{eqnarray}
All these expressions have the corrections of order $O(\varepsilon^{3})$
for $\eta$ and $O(\varepsilon^{2})$ for $1/\nu$.

For a single-charge model, the invariant coupling constant $\bar g$ always
lies in the interval $(0,g_{*})$ and necessarily tends to the IR attractive
fixed point $g_{*}$ as $s=k/\mu$ tends to zero. For a multicharge model,
even when an IR attractive fixed point is present, not every RG trajectory
(solution of the system (\ref{Odri})) will approach it for $s=k/\mu\to 0$.
A trajectory may first pass outside the region of stability (given by the
inequalities (\ref{stabR}) in our model), the situation usually interpreted
as a first-order phase transition. It may also go to infinity within the
stability region, the situation in which the perturbation theory becomes
unapplicable. Thus for every IR fixed point $g_{*}$ one can introduce the
notion of its basin of attraction: the set of all initial data $g$ for which
the solution of the system (\ref{Odri}) approaches $g_{*}$ in the limit
$s=k/\mu\to 0$.

\begin{figure}
\begin{center}
\includegraphics[width=15cm]{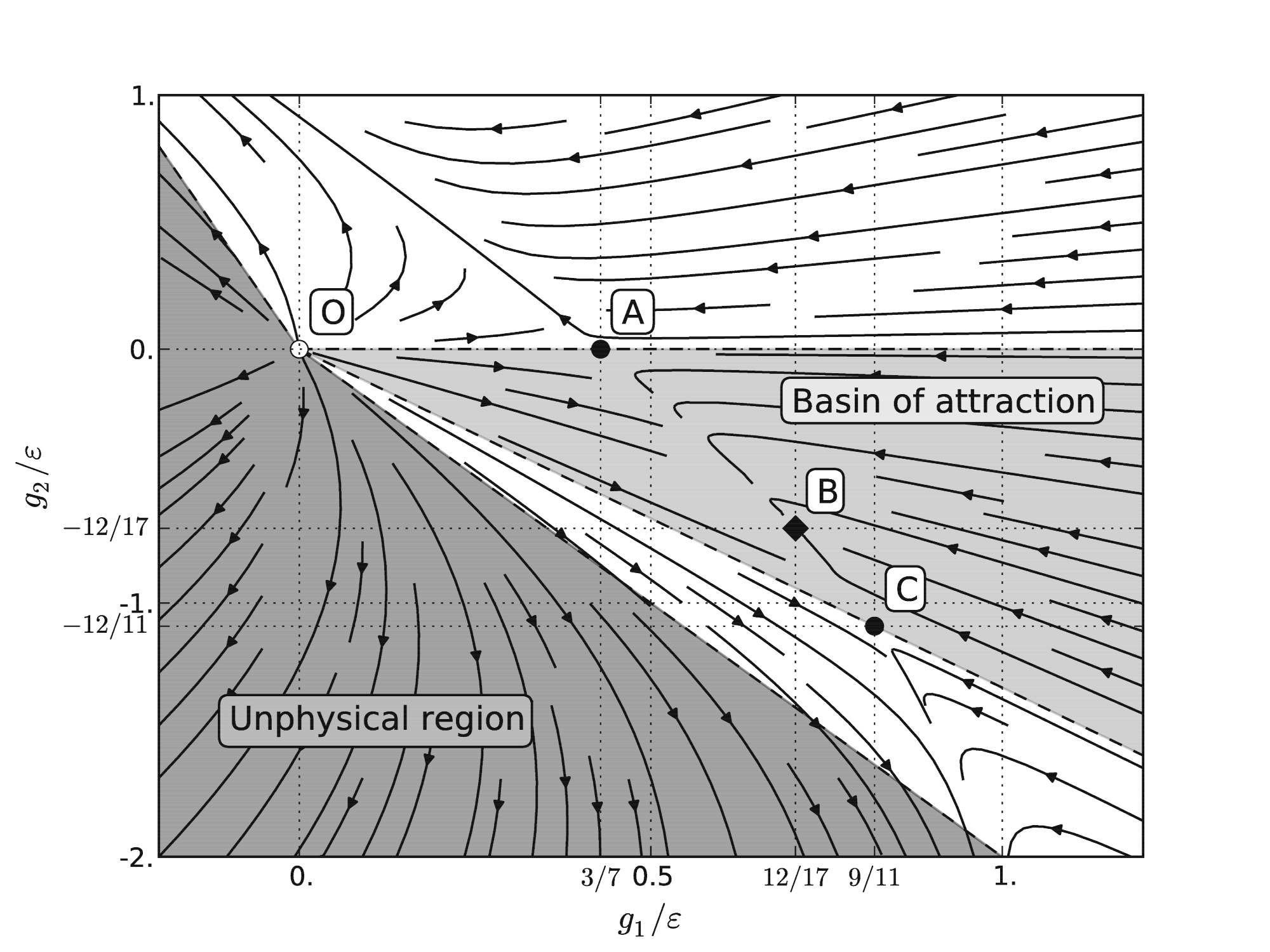}
\caption{\label{fig:1} RG flows for $n=4$.}
\end{center}
\end{figure}

The general pattern of the fixed points and RG flows in the $g_{1}$--$g_{2}$
plane for the case $n=4$ is shown on figure~1. In the one-loop approximation
it does not depend on $\varepsilon$ in the coordinates $g_{1,2}/\varepsilon$.
One can see that the IR attractive fixed point {\bf B} with the coordinates
(\ref{FP1}) lies between the two saddle points {\bf A} and {\bf C}, with the
coordinates (\ref{Fps}) and (\ref{FP2}), respectively. The origin houses the
IR repulsive Gaussian point {\bf O}.

Nevertheless, the RG flow approaches the point {\bf B} if the initial data
for the system (\ref{Odri}) lie in the vast basin of attraction (light grey).
Otherwise, the RG flow crosses the border of the stability region (given
by the inequality (\ref{stabR}) with $n=4$), thus coming into unphysical
region (dark grey).

It is worth noting that the basin of attraction lies entirely below the
$g_{2}=0$ axis. Indeed, for the initial data with $g_{2}>0$ (and thus
$g_{2}>-4g_{1}$ due to (\ref{stabR})) the RG flow has no chance to reach
the point {\bf B}: it cannot cross the axis $g_{2}=0$ because the
function $\beta_{2}$ vanishes there for all $g_{1}$:
$\beta_{2}({g_{2}=0})=0$, see expression (\ref{beta1}). Note that
$\beta_{1}({g_{1}=0}) \ne 0$, so that crossing the axis $g_{1}=0$
is allowed.

We may conclude that the IR behaviour in our model for $n=4$ is non-universal
in the sense that it depends on the choice of the couplings $g_{1,2}$.
If they belong to the basin of attraction for the IR point (in particular,
this implies $g_{2}<0$), the phase transition is of the second order (and
thus the scaling regime takes place). Otherwise the RG flows pass outside
the stability region. For $n>4$, there is no attractive fixed points and
only the second possibility can realize. This means that the account of
fluctuations changes the nature of the phase transition from the second-order
type (suggested by the mean-field theory) to the first-order type.

Surprisingly enough, the general situation is quite similar to the case of
complex antisymmetric order parameter \cite{NKH}, although the models are not
equivalent (in contrast to the real and complex vector cases). For $n>4$,
our RG flow is similar to that of the model with a third-rank tensor order
parameter, discussed in \cite{Leo1} in connection with the
isotropic-to-tetrahedratic transitions in liquid crystals.

In Ref. \cite{Vicari}, the following ``$\eta$ conjecture'' was formulated
for a general $\phi^{4}$-model with the vector order parameter: if the IR
attractive fixed point is present, it corresponds to the fastest decay of
correlations (that is, to the largest value of the exponent $\eta$). It is
easily checked that the expressions (\ref{indi14})--(\ref{indi3}) agree with
that conjecture for our tensor model (although the numerical values of
$\eta$ are very close to each other for all the three points). It should
be stressed that the proof given in \cite{Vicari} in the lowest order of the
$\varepsilon$-expansion does not apply to our case, because the
$\beta$-functions (\ref{beta1}) cannot be derived from a potential, that is,
they do not have the form $\beta_{i} = \partial U / \partial g_{i}$ with a
certain function $U=U(g_{1,2})$. Thus we have obtained an independent
confirmation of the $\eta$ conjecture for our special case of the tensor
model.

Of course, it is not impossible that all these results will change somehow
when
the higher-order contributions to the RG functions are taken into account;
cf. \cite{Sc2} for the electrodynamics of a charged scalar field. In order
to exclude (or to confirm) such a possibility, one has to calculate the RG
functions beyond the leading-order approximation and to apply an appropriate
resummation procedure. This work is already in progress.


\section*{Acknowledgments}

The authors are indebted to L.Ts. Adzhemyan, Michal Hnatich, Juha Honkonen,
M.Yu. Nalimov, P.V. Prudnikov, V.P. Romanov and A.I. Sokolov for helpful
discussion. The work was supported in part by the Russian Foundation for
Basic Research (project No 12-02-00874-a).

\end{document}